\documentclass[aps,10pt,notitlepage,nofootinbib,twocolumn]{revtex4-2}
\usepackage[utf8]{inputenc}
\usepackage{amsmath,amssymb,amsfonts}
\usepackage{newtxtext,newtxmath}
\usepackage{bbold}
\usepackage{bm}
\usepackage{graphicx}
\usepackage[usenames,dvipsnames]{xcolor}
\usepackage{color}
\usepackage[colorlinks=true,linkcolor=blue,urlcolor=blue,citecolor=blue]{hyperref}

\usepackage{slashed}
\usepackage[english]{babel}
\usepackage{dcolumn}
\usepackage{pifont}
\usepackage{dsfont,mathrsfs}
\usepackage{cancel}
\usepackage{bigints}

\usepackage{lineno}
\usepackage{amsfonts}
\usepackage{amsmath}
\usepackage{amssymb}
\usepackage{graphicx,color}
\usepackage{float}
\usepackage{hyperref}
\usepackage{subfigure}
\usepackage{mathtools,slashed}
\usepackage{ulem}
\usepackage{lipsum}

\def\nn {\nonumber}

\def\be {\begin{equation}}
\def\ee {\end{equation}}
\def\bea {\begin{eqnarray}}
\def\eea {\end{eqnarray}}
\def\bc {\begin{center}}
\def\ec {\end{center}}

\begin{document}

\title{Estimation of the  diffusion coefficient of Heavy Quarks  in light of Gribov-Zwanziger action}

\author{Sadaf Madni$^{a}$}
\email{sadaf.madni@niser.ac.in}
\author{Arghya Mukherjee$^{a,b}$}
\email{arbp.phy@gmail.com}
\author{Aritra Bandyopadhyay$^{c}$}
\email{ aritrabanerjee.444@gmail.com}
\author{Najmul Haque$^{a}$}
\email{nhaque@niser.ac.in}

\affiliation{$^a$School of Physical Sciences, National Institute of Science Education and Research, HBNI, Jatni, Khurda 752050, India}
\affiliation{$^b$Department of Physics and Astronomy,Brandon University, Brandon, Manitoba R7A 6A9, Canada}
\affiliation{$^c$Institut f\"ur Theoretische Physik, Universit\"at Heidelberg, Philosophenweg 16, 69120 Heidelberg, Germany}

\begin{abstract}
 The heavy quark momentum diffusion coefficient ($\kappa$) is one of the most essential ingredients for the Langevin description of heavy quark dynamics. In the temperature regime relevant to the heavy ion collision phenomenology, a substantial difference exists between the lattice estimations of $\kappa$ and the corresponding leading order (LO) result from the hard thermal loop (HTL) perturbation theory. Moreover, the indication of poor convergence in the  next-to-leading order (NLO) perturbative analysis has motivated the development of several approaches to incorporate the non-perturbative effects in the heavy quark phenomenology. In this work, we  estimate the heavy quark diffusion coefficient  based on the Gribov-Zwanziger prescription. In this framework, the gluon propagator depends on the temperature-dependent Gribov mass parameter, which has been obtained self-consistently from the one-loop gap equation. Incorporating this modified gluon propagator in the  analysis, we find a reasonable agreement  with the existing lattice estimations of $\kappa$ within the model uncertainties. 
\end{abstract}

\maketitle

\medskip

\section{Introduction}
 Since the seminal work by Svetitsky~\cite{Svetitsky:1987gq}, there has been a concerted effort to understand the heavy quark dynamics in the presence of the  evolving background medium  produced in the relativistic heavy-ion collision experiments ~\cite{Andronic:2015wma,Rapp:2018qla,Dong:2019byy,Zhao:2020jqu}.  An intriguing feature associated with the  heavy quarks is the inherent distinction of their dynamics compared to the light quark sector. With orders of magnitude differences in their masses, the heavy quarks can only achieve a slower equilibration rate relative to their lighter counterparts, rendering their thermalization time comparable to or even larger than  the lifetime of the fireball. As a consequence, an imprint of  the interaction history is expected to be retained in the final state  heavy quark spectra \cite{Prino:2016cni}, which makes them an important probe for the characterization of the strongly interacting background. On the other hand,  the typical momenta carried by the heavy quarks being much larger than the ambient temperature scale,  a single collision with the medium constituents can hardly change the momentum significantly. Hence, a Brownian motion with successive uncorrelated momentum kicks  has been a well-accepted description of the dynamics of the heavy quarks. The corresponding Langevin equation governing the heavy quarks' momentum evolution possesses a drag force whose  strength  is determined by  the associated drag coefficient ($\eta_D$). Whereas the random momentum kicks are incorporated in the momentum evolution through a stochastic force term $\xi_i(t)$ having a correlation $\langle \xi_i(t)\xi_j(t')\rangle$ proportional to $\delta_{ij}\delta(t-t')$. The proportionality constant is related to the mean squared momentum transfer per unit time and is known as the momentum diffusion coefficient $\kappa$~\cite{Moore:2004tg}.  In equilibrium,  the drag coefficient is related to the momentum diffusion coefficient  through the Einstein relation $\eta_D=\kappa/(2 M T)$ where $M$ corresponds to the mass of the propagating heavy quark and $T$ is the temperature of the medium. At vanishing momentum transfer, the two coefficients can further be related to the spatial diffusion coefficient $D_s$  as $D_s=T/(M \eta_D)=2 T^2/\kappa$. This simplification allows one to conveniently encode the medium characteristics in a single transport coefficient conventionally defined as $(2\pi T)D_s$, where the spatial diffusion coefficient is scaled with the thermal wavelength.  Within the perturbative approach, the heavy quark diffusion coefficient has been obtained in the LO~\cite{Svetitsky:1987gq,Moore:2004tg} as well as in the NLO~\cite{Caron-Huot:2007rwy} with the strong coupling. From those studies, it has been observed that for realistic coupling strength, the NLO corrections become  several times larger than the LO estimates, reflecting the poor convergence of the perturbation series. Hence a non-perturbative estimation is crucial to constrain the temperature dependence of the diffusion coefficient, which has important  implications in the heavy ion phenomenology~\cite{Das:2015ana}. Since the first attempt~\cite{Banerjee:2011ra} based on the definition  obtained in Ref.~\cite{Caron-Huot:2009ncn}, there have been several non-perturbative estimations for the  diffusion coefficient using the lattice discretized QCD approach (see for example Refs. \cite{Francis:2015daa,Brambilla:2020siz,Altenkort:2020fgs,Brambilla:2022xbd,Banerjee:2022gen} ) which significantly enriched our knowledge of the temperature dependence of the diffusion coefficient. Though large uncertainties exist in the lattice estimates, their consistency within the error reconfirms the poor convergence of perturbative estimation. Although it should be noted here that these non-perturbative estimations of $\kappa$ are restricted to the quenched approximation  in the  lattice framework,  which assumes a gluonic plasma  and hence does not represent the realistic scenario expected in the heavy ion collision experiments. However, the quenched approximated results are of similar magnitude as expected in the temperature regime relevant in the phenomenology (see, for example, Refs.~\cite{Rapp:2018qla,Cao:2018ews} for a comprehensive comparison between the lattice results and the estimations  from different phenomenological models as well as data-driven approaches \cite{Xu:2017obm}). At this point, it is interesting to note that the estimation of  the heavy quark transport coefficient in the phenomenologically relevant  regime of temperatures  is not the only scenario where the perturbative estimation differs from the lattice results obtained with pure gluonic plasma. For example, in the temperature regime close to $T_c\sim270$ MeV, a significant deviation can be observed when the free energy data obtained from the pure Yang-Mills lattice simulation \cite{Borsanyi:2012ve} is compared with  the  resummed perturbative theory estimates~\cite{Andersen:2009tc}. In this context, a remarkable improvement over the resummed perturbative approach has been achieved in Ref.~\cite{Fukushima:2013xsa} implementing the Gribov-Zwanziger (GZ) prescription~\cite{Gribov:1977wm,Zwanziger:1989mf} in the evaluation of the free energy.

The Gribov-Zwanziger formalism \cite{Dokshitzer:2004ie,Maas:2011se} improves the infrared behaviour of QCD by fixing the residual gauge transformations that remain after the implementation of the Faddeev-Popov quantization procedure. A review on the Gribov prescription can be found in Refs.~\cite{Sobreiro:2005ec,Vandersickel:2012tz} (for recent progress  on the refined Gribov approach, see also 
\cite{Dudal:2008sp,Capri:2016aqq,Dudal:2017kxb,Gotsman:2020ryd,Gotsman:2020mpg,Justo:2022vwa} and the references therein). Within the GZ framework, the gluon propagator in the  Landau gauge is expressed as~\cite{Fukushima:2013xsa}
\bea
G^{\mu\alpha}(Q) = \left[\delta^{\mu\alpha}-\frac{Q^\mu Q^\alpha}{Q^2}\right]\frac{Q^2}{Q^4+\gamma_G^4},\label{t_gp1}
\eea
where  $\gamma_G$ is the Gribov parameter. This parameter in the denominator shifts the poles of the gluon propagator off the energy axis to an unphysical location $Q^2 = \pm i\gamma_G^2$, suggesting that the gluons are not physical excitations. In the context of  finite temperature Yang-Mills (YM) theory, the  temperature dependence of the Gribov parameter can be determined by solving a gap equation self-consistently. As a result, one obtains a non-perturbative IR cut-off in the gluon propagator, which is absent in the traditional YM approach where the finite temperature perturbative expansion breaks down at the (chromo)magnetic scale $g^2 T$, known as the Linde problem~\cite{Linde:1980ts}. It was first pointed out in Ref.~\cite{Zwanziger:2006sc} that, in the limit of asymptotically high temperatures, $\gamma_G$ becomes proportional precisely to the (chromo)magnetic scale $g^2 T$  indicating a promising resolution to such IR-catastrophe. Recently, the Gribov modified gluon propagator with this asymptotic temperature dependence has been implemented in the hard-thermal-loop (HTL) effective theory to study the quark dispersion~\cite{Su:2014rma} relations, the dilepton rate, and the quark number susceptibility~\cite{Bandyopadhyay:2015wua}, which revealed some intriguing physical properties of the said observables.  
Thus, it is interesting to investigate whether the effects from the  magnetic scale incorporated via the Gribov-Zwanziger prescription can improve the leading order  perturbative estimation of the heavy quark diffusion coefficient in the phenomenologically relevant temperature regime. In the present work, we address this issue considering two scenarios: first, we consider pure gluonic plasma as the background medium for the heavy quarks. In this case, the temperature dependence of the Gribov parameter is extracted using the one-loop gap equation.  Next, we consider the heavy quarks in the QGP background with only the  asymptotic temperature dependence of $\gamma_G$. It turns out that the corresponding momentum diffusion coefficient in both scenarios shows significantly different temperature dependence  compared to the leading order perturbative result.

 The letter is organized as follows. At first in section~\ref{formalism}, 
 the momentum diffusion coefficient is obtained from  the elastic scattering amplitude in the non-relativistic limit implementing the Gribov-modified gluon propagator. The obtained diffusion coefficient $\kappa$ depends explicitly on the temperature-dependent Gribov parameter $\gamma_G(T)$, which is subsequently determined in section~\ref{gap}. Next, in section~\ref{results}, the numerical estimation of the  scaled diffusion coefficient is compared with the available lattice data  and the perturbative results. Finally, we summarize the main results in section~\ref{summary} and conclude with a brief  discussion of the possible future directions.

\medskip


\medskip

\section{Formalism}\label{formalism}
The Langevin equation corresponding to the momentum evolution of  the non-relativistic heavy quarks in a background medium is given by~\cite{Moore:2004tg}
\begin{equation}
\frac{d p_i }{dt}=-\eta_D p_i +\xi_i(t)\,,\hspace{0.3cm}\langle\xi_i(t)\xi_j(t')\rangle=\kappa \delta_{ij}\delta(t-t')\,.
\end{equation}
Here the momentum diffusion coefficient $\kappa$ can be obtained from the mean squared momentum transfer per unit time, which is  $3\kappa$ and which can be computed from the expression~\cite{Moore:2004tg}:
\begin{align}
3\kappa &= \frac{1}{2M}\int\frac{d^3{\bm k}d^3{\bm k^\prime}d^3{\bm p^\prime}}{(2\pi)^9(2k^0)(2k^{\prime 0})(2M)}   \nn\\
&\times  ({\bm k- \bm k^\prime })^2 (2\pi)^4\delta^4(P^\prime+K^\prime-P-K)\nn\\&\times\left\{|\mathcal{M}|^2_{\rm quark}\, n_F(k)[1-n_F(k^\prime)] \right. \nn\\
&\left. +\ |\mathcal{M}|^2_{\rm gluon}\, n_B(k)[1+n_B(k^\prime)]\right\}\,,
\label{interaction_rate1}
\end{align}
where $|\mathcal{M}|^2_{\rm quark}$ and $|\mathcal{M}|^2_{\rm gluon}$ represent the squared $2\leftrightarrow 2$ scattering matrix elements  corresponding to $qH\rightarrow qH$ and $gH\rightarrow gH$ processes respectively. Note that, here,  the  $q$ and $g$ respectively  correspond to the incoming  or outgoing massless light quarks (as well as the anti-quarks), and  gluons of the background medium with four-momentum denoted as $K\equiv(k^0,{\bm k})$ or $K^\prime$. Unlike the heavy quarks (represented by $H$), these medium constituents  possess the corresponding statistical weight factors, which are  expressed in terms of  the  Fermi-Dirac ($n_F$) and the Bose-Einstein ($n_B$) distribution functions, respectively, which, in the massless limit,  depending on the magnitude of the three momentum vectors denoted as $k$ or $k^\prime$. 
Moreover, in the above expression, the heavy quarks are considered non-relativistic, which essentially makes the energies of the incoming ($P$) and the outgoing ($P^\prime$) heavy quarks identical  to their rest mass energy $M$. Now, in the leading order, the squared scattering matrix element for the $qH\rightarrow qH$  process  via the $t$-channel gluon exchange, together with the anti-quark contribution, can be expressed as 
\begin{align}
	|\mathcal{M}|^2_{\rm quark} = N_f C_F g^4 L_{\mu\nu}M_{\alpha\beta} G^{\mu\alpha}(Q)G^{\nu\beta}(Q)\,,
	\label{ssme_q}
\end{align}
where  $C_F$ is the color Casimir constant, $G^{\mu\alpha}$ is the  gluon propagator that depends on the momentum transfer $Q=P^\prime-P=K-K^\prime$. Note that the squared matrix amplitude has  been summed and averaged over the final and the initial  quantum numbers (spins and colors), respectively, and multiplied with the degeneracy factor of the initial light quark ($2N_c$)  and  the number of flavors ($N_f$) along with a factor of 2 for the anti-quark contribution.  The Dirac traces over  the  quark external lines along with the metric $g^{\mu\alpha}$ and $g^{\nu\beta}$ in the gluon propagators provide 
\begin{align}
	L_{\mu\nu}M^{\mu\nu}&=16 \big[(P\cdot K)(P^\prime\cdot K^\prime)\nn\\&+(P\cdot K^\prime)(P^\prime\cdot K)- M^2  (K\cdot K^\prime)\big]\,. \label{tr}
\end{align}
Now, considering the heavy quark  approximately at rest, i.e. $P^\mu = (M,0)$,  $Q$  becomes purely spatial, and we have, ${\bm k^\prime} = {\bm k} - {\bm q}$, which gives the following simplified relations between the momenta: 
\begin{align}
&\hspace{-0.4cm}K \cdot K^\prime = k_0^2-{\bm k}\cdot{\bm k^\prime} = k^2(1-\cos\theta_{\bm{ kk^\prime}})\,, \label{mom_frame1}\\
&\hspace{-0.4cm}(P\cdot K)(P^\prime\cdot K^\prime)+(P\cdot K^\prime)(P^\prime\cdot K) = 2M^2k^2\,.\label{mom_frame2}
\end{align}
Note that, as the heavy quarks are considered to be approximately at rest, we essentially need to incorporate only the temporal part of the gauge boson propagator, i.e., $G^{00}(Q)$ in the squared matrix element. Now, in the standard approach, to incorporate the  screening in the  t-channel  amplitude, the free gluon propagator is replaced with the  HTL resummed one loop effective propagator, which, in the limit of small energy transfer, is  equivalent to introducing the Debye mass ($m_D$) in the denominator~\cite{Moore:2004tg}. However, in the present case, we consider the $G^{00}(Q)$ component of the modified gluon propagator within GZ approach, which, along with the fact that the energy transfer is negligible (i.e., $Q$ is purely spatial), gives   
\begin{align}
G^{\mu\alpha}(Q) = -\frac{\delta^{\mu 0}\delta^{\alpha 0}q^2}{q^4+\gamma_G^4}\,.
\label{t_gp_simple}
\end{align}
Incorporating this modified propagator in  Eq.~\eqref{ssme_q} along with the simplified relationships given in  Eq.~\eqref{mom_frame1} and \eqref{mom_frame2}, we obtain the final expression for the squared  matrix element for the $qH\rightarrow qH$ channel as 
\begin{align}
|\mathcal{M}|^2_{\rm quark} &= 16N_fC_Fg^4 M^2 \frac{k^2 q^4 (1+\cos\theta_{\bm kk^\prime})}{(q^4+\gamma_G^4)^2}\,.
\label{ssme_q_final}
\end{align}

The scattering matrix  for  the $gH\rightarrow gH$ process involves the three gluon vertex $\Gamma_{\rho\sigma\alpha}(K,K^\prime,Q)$ in the t-channel. The complete expression  of the squared matrix element considering the s, t, and u channel processes can be found in Refs.~\cite{Combridge:1978kx,Berrehrah:2013mua}. As argued in \cite{Moore:2004tg}, in the rest frame of the heavy quark, the matrix element is dominated by the t-channel gluon exchange, and the  Compton amplitude can be safely ignored. Incorporating the modified gluon propagator in the t-channel, we obtain
\be
|\mathcal{M}|^2_{\rm gluon} = 16N_cC_Fg^4 M^2\frac{k^2 q^4 (1+\cos^2\theta_{\bm{kk^\prime}})}{(q^4+\gamma_G^4)^2}\,,
\label{ssme_g_final}
\ee
where an additional degeneracy factor $2(N_c^2-1)$ corresponding to the initial gluon has been multiplied with the spin and color-averaged (summed) squared amplitude.  

Now, the obtained squared matrix elements from Eqs.~\eqref{ssme_q_final} and~\eqref{ssme_g_final} can be inserted within the integrals  given in Eq.~\eqref{interaction_rate1}. It is convenient to shift the integral over ${\bm p^\prime}$ to an integral over the momentum transfer $\bm{q}=\bm{p-p^\prime}$ and express the diffusion coefficient as 
\begin{align}
3\kappa &= \frac{1}{2M} \int\frac{d^3{\bm k}d^3{\bm k^\prime}d^3{\bm q}}{(2\pi)^98kk^\prime M}(2\pi)^3\delta^3({\bm k^\prime}-{\bm q}-{\bm k})\nn\\
&\times~2\pi \delta(k^\prime - k)q^2\times  \Bigl[|\mathcal{M}|^2_{\rm quark}n_F(k)[1-n_F(k^\prime)] \nn\\
& + |\mathcal{M}|^2_{\rm gluon}n_B(k)[1+n_B(k^\prime)]\Bigr]\,.
\end{align}
Here we note that  the squared matrix amplitudes depend only on the magnitude of the integration variables ${\bm q}$ and ${\bm k}$ as their angular dependence can be re-expressed using 
\begin{align}
 \cos\theta_{\bm{kk^\prime}}&=1-\frac{q^2}{2k^2}\,.
\end{align}
Thus, the ${\bm k^\prime}$ integral over the three-momentum delta function can be easily performed, and we obtain 
\begin{align}
3\kappa &= \frac{1}{16M^2} \int\frac{d^3{\bm k}}{(2\pi)^4k}\int q^2dq \int\limits_{-1}^1 d\cos\theta_{{\bm{ kq}}}\nn\\
&\times\Big[q~\delta(k^\prime - k)~\frac{q}{k^\prime} \Big\{|\mathcal{M}|^2_{\rm quark}n_F(k)[1-n_F(k^\prime)] \nn\\
& +\ |\mathcal{M}|^2_{\rm gluon}n_B(k)[1+n_B(k^\prime)]\Big\}\Big]_{\bm{ k^\prime=k+q}}\,.
\end{align}
Next, the energy delta function can be integrated over the variable $\cos\theta_{\bm{kq}}$ by using the property  
\begin{align}
q~\delta(k^\prime -k)& =  \delta\left(\cos\theta_{\bm{kq}}+\frac{q}{2k}\right)\,,
\end{align}
which also restricts the upper limit of the $q$ integral to $q_{\rm max}=2k$. After this angular integral, we obtain 
\begin{align}
3\kappa &= \frac{C_Fg^4}{4\pi^3} \int\limits_0^\infty k^2dk\int\limits_0^{2k} qdq\, \chi(q,\gamma_G) \nn\\
&\times\left[N_fn_F(k)[1-n_F(k)]\left(2-\frac{q^2}{2k^2}\right) \right.\nn\\
&\left.+ N_c n_B(k)[1+n_B(k)]\left(2-\frac{q^2}{k^2}+\frac{q^4}{4k^4}\right)\right]\,,\label{eqn:1}
\end{align}
where the $\gamma_G$ dependence only appears in the factor
\begin{align}
 \chi(q,\gamma_G)&=\frac{q^6}{\left(q^4+\gamma_G^4\right)^2}\,,
\end{align}
which we have introduced for our convenience.
One can further integrate the $q$ dependence, which yields
\begin{align}
&\hspace{-0.5cm}3\kappa = \frac{C_Fg^4 T^3}{4\pi^3} \int\limits_0^\infty x^2dx \Biggl[N_fn_F(x)[1-n_F(x)]\nn\\
&\hspace{-0.2cm}\times\left\{\frac{3\bar{\gamma}_G^2}{8x^2}\arctan\frac{4x^2}{\bar{\gamma}_G^2} +\frac{1}{2}\ln\frac{16x^4+\bar{\gamma}_G^4}{\bar{\gamma}_G^4}-\frac{3}{2}\right\} \nn\\
&\hspace{-0.2cm}+ N_c n_B(x)[1+n_B(x)]\Biggl\{\frac{3\bar{\gamma}_G^2}{4x^2}\arctan\frac{4x^2}{\bar{\gamma}_G^2} -1\nn\\
&\hspace{-0.2cm} +\left(\frac{1}{2}-\frac{\bar{\gamma}_G^4}{8x^4}\right)\ln\frac{16x^4+\bar{\gamma}_G^4}{\bar{\gamma}_G^4}-\frac{8x^4}{16x^4+\bar{\gamma}_G^4}\Biggr\}\Biggr]\,,\label{q-integrated}
\end{align}
where $\bar{\gamma}_G=\gamma_G(T)/T$ and the integral is represented in terms of the dimensionless variable $x=k/T$. Note that, similar to the Debye mass, the Gribov parameter itself is a function of temperature, and it  plays a crucial role in determining the  overall temperature dependence of the diffusion coefficient, as we will see later. Let us first obtain the diffusion coefficient in the limit of small $\bar{\gamma}_G$. In that case, it can be  compared with the standard $m_D/T$ expanded results, which, including the NLO corrections, are given by~\cite{Caron-Huot:2007rwy}  
\begin{align}
 &\hspace{-0.8cm}3\kappa=\frac{C_F g^4T^3}{6\pi}\left[\Big(N_c+\frac{N_f}{2}\Big)\Big[\ln\frac{2T}{m_D}+\frac{1}{2}-\gamma_E+\frac{\zeta^\prime(2)}{\zeta(2)}\Big]\right.\nn\\&\hspace{1.6cm}\left.+\frac{N_f}{2}\ln2+ 2.3302~ \Big(N_c \frac{m_D}{T}\Big) \right]\,,\label{NLO}
\end{align}
 where the NLO correction is incorporated in the last term. The LO result in Eq.~\eqref{NLO} can be obtained from Eq. \eqref{eqn:1} by simply replacing the factor $\chi(q,\gamma_G)$ with the Debye screened expression $\chi(q, m_D)$ given by 
 \begin{align}
 \chi(q,m_D)&=\frac{q^2}{\left(q^2+m_D^2\right)^2}\,,
\end{align}
and performing the subsequent $q$ and $k$ integration. We can further establish the correlation between the LO Debye screened case explored earlier \cite{Moore:2004tg} and our results by denoting the LO expansions as  $(3\kappa)^{\rm LO}$  and $(3\kappa)_{\gamma_G}$ respectively for the Debye screened case and the Gribov case. A straightforward calculation shows :
\begin{align}
 (3\kappa)_{\gamma_G}&=(3\kappa)^{\rm LO}_{m_D\rightarrow\gamma_G}+C_F m_D^2\frac{ g^2 T}{8\pi}\,,
\end{align}
where,  the perturbative definition of $m_D^2=(1/3)g^2T^2(N_c+N_f/2)$ is used in the last term, which essentially arises from the difference in the momentum dependence of the  gluon propagator compared to the  Debye screened propagator in the standard approach. At asymptotically high temperatures, this  additional contribution  becomes ${\sim}\gamma_G m_D^2$ (as $\gamma_G$ corresponds to the (chromo)magnetic scale $g^2T$) whereas the leading log dependence i.e. ${\sim}g^4\log(T/m_D)$ now becomes ${\sim}g^4\log(T/\gamma_G)$, thus, remains ${\sim}g^4\log(1/g)$. However, for phenomenologically relevant temperatures, one may require a non-perturbative estimation of the temperature dependence of the Gribov parameter,  as will be  discussed in the following.      

\section{Fixing the Gribov parameter}\label{gap}
\begin{figure}[tbh]
	\center
	\includegraphics[width=\linewidth]{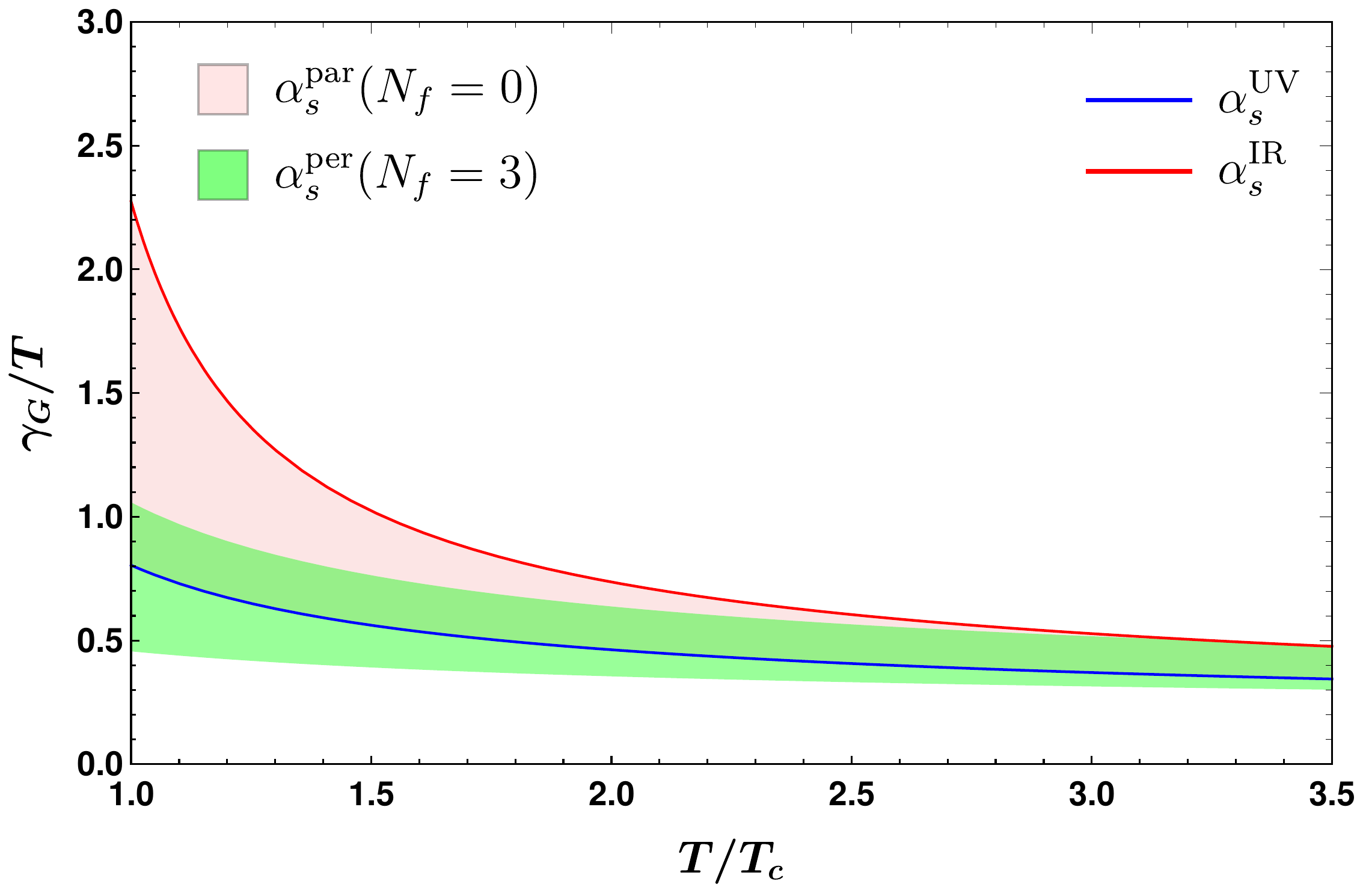}
	\caption{The temperature dependence of the scaled Gribov parameter with $N_f=0$ is shown along with the $N_f=3$ case obtained considering the asymptotic form and the 1-loop perturbative running. The bands refer to the variation of the scale  from $w=w_{\rm IR}$  to $w_{\rm UV}$  and $w=1$ to 4 respectively. }
	\label{gammaG_T_dependence}
\end{figure}

As mentioned earlier, the Gribov prescription improves the Faddeev-Popov quantization procedure by modifying the YM  partition function in the Euclidean space-time as~\cite{Vandersickel:2012tz}
\begin{align}
	Z&=\int [\mathcal{D}A][\mathcal{D} \Bar{c}][\mathcal{D}c]\,V(\Omega)\delta(\partial_{\mu}A_{\mu}^a)\nn\\
	&\times \exp\Big[-S_{\rm YM}-\!\int\!\!\! d^4x \Bar{c}^a(x)\partial_{\mu}D_{\mu}^{ab}c^b(x)\Big]\,,
	\label{Z}
	\end{align}
	where, $D_{\mu}^{ab}$ is the covariant derivative, $c \hspace{0.1
	cm}{\rm and }\hspace{0.1cm} \Bar{c}$ are the ghost, and anti-ghost fields, and the no pole condition is implemented by the step function $V(\Omega)  = \theta (1-\sigma_0) $ that  constraints the path integral over the field configurations  within the Gribov region $\Omega$  defined  as 
	\begin{align}
	\Omega\equiv\{A_{\mu}^a,\partial_{\mu}A_{\mu}^a=0 | -\partial_{\mu}D_{\mu}^{ab}>0 \}\,.
\end{align}
 The function $\sigma_0=\sigma(P=0)$  appears in the inverse of the ghost dressing function as $Z_G^{-1}=1-\sigma(P)$  corresponding to the ghost field propagator given by
 \begin{align}
	D_c(P)&= \frac{\delta^{ab}}{1-\sigma(P)} \frac{1}{P^2}\,.
\end{align}
 The standard procedure to obtain the momentum space gluon propagator $\langle A_\mu^a(P)A_\nu^b(K)\rangle$ from the  path integral as given in Eq.~\eqref{Z}, is to re-express the constraint  as
\begin{align}
  V(\Omega)&=\frac{1}{2\pi i}\int_{-i\infty+0^+}^{i\infty+0^+} \frac{ds}{ s}\exp\big[s(1-\sigma_0)\big]\,,
\end{align}
and consider the steepest descent approximation for the integral over the variable $s$. Introducing a mass parameter $\gamma_G$  corresponding to the  specific value $s=s_0$ that minimizes the entire exponent of the integral, one obtains  the gap equation for the Gribov parameter (see Ref.~\cite{Vandersickel:2012tz} for details).  In the case of finite temperature,  the  gap equation  is given by   
\begin{align}
    \sumint_P\frac{1}{P^4+\gamma_G^4}&=\frac{d}{(d-1)N_c g^2}\,,
\end{align}	
where $d$ is the space-time dimension, and the temporal component of the Euclidean four-momentum $P=(\bm{p},p_4)$ possesses the discreet Matsubara frequencies $p_4=2 n \pi T$. It is straightforward to perform the frequency sum and evaluate the momentum integral using dimensional regularization. In the $\overline{\rm MS}$
renormalization scheme with $d=4$, one obtains~\cite{Gracey:2005cx}
\begin{align}
\frac{3 N_c g^2}{64 \pi^2}\Bigg[\frac{5}{6}-&\ln\frac{\gamma_G^2}{\mu_0^2}+\frac{4}{i \gamma_G^2}\displaystyle\int_{0}^{\infty}dp p^2\nn\\
	&\times \left(\frac{n_B(\omega_-)}{\omega_-}-\frac{n_B(\omega_+)}{\omega_+}\right)\!\!\Bigg]=1,\label{gap_eqn} \end{align}
where $\mu_0$ represents the regularization scale and the argument in the distribution function
$n_B(\omega)=(e^{\omega/T}-1)^{-1}$ corresponds to the dispersion relations, which in this case are, $\omega_{\pm}=\sqrt{p^2 \pm i \gamma_G^2}$. On the other hand, the  sum integral corresponding to the $\sigma(P)$ in the ghost dressing function is given by~\cite{Fukushima:2013xsa},
\begin{align}
	&\sigma(P)=N_c g^2 \frac{P_{\mu} P_{\nu}}{P^2}\nn\\ &\times\displaystyle\sumint_{Q}\frac{1}{Q^4+\gamma_G^4}\frac{Q^2}{(Q-P)^2} \left(\delta^{\mu\nu}-\frac{Q^{\mu}Q^{\nu}}{Q^2}\right)~. \label{sigma}
\end{align}
To obtain the temperature dependence of the Gribov parameter from the gap equation given in Eq.~\eqref{gap_eqn}, it is essential first to fix the scale $\mu_0$ that appears in the temperature-independent part. Following Ref.~\cite{Fukushima:2013xsa} for this purpose, we first consider the $T=0$ contribution of the gap equation given by 
\begin{align}
    	\gamma_{G0}&=\mu_0\hspace{0.1cm} \exp \left(\frac{5}{12}-\frac{32 \pi^2}{3 N_c g_0^2}\right)\,,
\end{align}
with $g_0$ as the strong coupling at vanishing temperature. The above expression of $\gamma_{G0}$ is then substituted   in the temperature-independent contribution of the inverse ghost dressing function, which can be obtained from Eq.~\eqref{sigma} using dimensional regularization  as
 \begin{align}
&\hspace{-0.7cm}1-\sigma(P)=\frac{N_c g_0^2}{128 \pi^2}\left[-5+\left(3-\frac{\gamma_{G0}^4}{P^4}\right)\log\left(1+\frac{P^4}{\gamma_{G0}^4}\right)\right.\nn\\
	&\hspace{0.5cm}\left.+\ \frac{\pi P^2}{\gamma_{G0}^2}+2\left(3-\frac{P^4}{\gamma_{G0}^4}\right)\frac{\gamma_{G0}^2}{P^2}{\rm arctan}\frac{P^2}{\gamma_{G0}^2}\right]\,,\label{ghostdf}
\end{align}
which provides the $\mu_0$ as a function of the coupling strength $g_0$. Considering $g_0=3.13$, one obtains $\mu_0=1.69$ GeV, which will be used in the finite temperature gap equation along with the  temperature-dependent running coupling $g(T)$. 
In Ref.~\cite{Fukushima:2013xsa}, the authors have parameterized the non-perturbative running coupling with a single parameter (say $w$) by  utilizing the functional form of the  one-loop perturbative coupling with $N_f=0$. It is shown that this simple, functional form.   
\begin{align}
\alpha^{\rm par}_s(T/T_c)&= \frac{6 \pi}{11  N_c \log [w\,(T/T_c)]}~,\label{coupling}
\end{align}   
 can  fit  the lattice-measured coupling data~\cite{Kaczmarek:2004gv} reasonably well for a wide range of temperatures. The fitted parameter values corresponding to the coupling data extracted from the large distance (IR) and the short distance (UV) behaviour of the heavy quark free energy  are given by $w_{\rm IR}=1.43$ and $w_{\rm UV}=2.97$ respectively. 
  In the present work, we consider this parametrized running coupling $g^2(T/T_c)=4\pi\alpha^{\rm par}_s(T/T_c)$ in the gap equation to numerically extract the temperature dependence  of the Gribov parameter for the  gluonic plasma ($N_f=0$). The obtained temperature dependence of the scaled Gribov parameter corresponding to the IR and the UV parametrization is shown in Fig.~\ref{gammaG_T_dependence}~\cite{Fukushima:2013xsa}. The uncertainty band, in this case, corresponds to the variation of the parameter $w$ from $w_{\rm IR}$ to  $w_{\rm UV}$. 
  Note that the temperature dependence of the scaled Gribov parameter in the IR case is similar to the decreasing trend obtained using the kinetic theory framework considering the quasiparticle picture of the Gribov plasma~\cite{Florkowski:2015dmm,Florkowski:2015rua,Jaiswal:2020qmj}. However, unlike the IR scenario, the obtained $\bar{\gamma}_G$ for $w=w_{\rm UV}$ remains small throughout the considered temperature range.  Moreover, as discussed in  Ref.~\cite{Fukushima:2013xsa}, the fitted coupling, along with the obtained $\mu_0$, provide an almost temperature-independent  ghost dressing function (as obtained  from Eq.~\eqref{sigma} numerically)  which is consistent with the lattice results~\cite{Cucchieri:2007ta,Aouane:2011fv}. However, it should be mentioned that in the original GZ approach, the dressing function shows an IR-enhancement at $P\rightarrow 0$ whereas the recent lattice measurements at vanishing momentum support a finite decoupling behaviour  which is in better agreement with the refined GZ approach~\cite{Dudal:2008sp,Capri:2016aqq}. Nevertheless, we adopt the original GZ procedure here due to its  considerable simplicity compared to the refined GZ approach, especially for finite temperature applications.
 
It is also interesting to compare the temperature dependence of the Gribov-modified diffusion coefficient with the perturbative result, including the quark contribution. For that purpose, we utilize the asymptotic temperature dependence of the Gribov parameter that can be obtained from the gap equation considering the  $T \rightarrow \infty$ limit as 
\begin{align}
	\gamma^{\rm asy}_G(T)&=\frac{d-1}{d}\frac{N_c}{4\sqrt{2}\pi}g^2(T)T\,. \label{gammaG_asym}
\end{align}
Thus, the temperature dependence of $\bar{\gamma}_G$, in this case, is completely determined by the running coupling. Keeping in mind the functional form of the parametrized coupling for $N_f=0$, in this case, we use the one-loop running coupling with $N_f=3$, which is given by
\begin{align}
    \alpha^{\rm per}_s(T)&= \frac{6 \pi}{(11  N_c -2 N_f) \log \left[w ( \pi T)/\Lambda_{\rm \overline{MS}}\right]}~,\label{per_coupling}
\end{align}
where, $\Lambda_{\rm \overline{MS}}=0.176$ GeV is  obtained from the  lattice measurement  $\alpha_s(1.5 {\rm GeV},N_f=3)=0.326$~\cite{Bazavov:2012ka} and  we consider  $T_c=0.16$ GeV to express the running coupling in terms of the scaled temperature $T/T_c$. The uncertainty band, in this case, is obtained by varying the energy scale symmetrically around $2\pi T$ by a factor of two (\textit{i.e.} $1\le w\le 4$). As can be observed from Fig.~\ref{gammaG_T_dependence},  with $N_f=3$, where the asymptotic form of the Gribov parameter is considered, the temperature dependence becomes similar to the UV case obtained previously with $N_f=0$. In the following, we consider the two scenarios with $N_f=0$ and three and discuss the influence of the temperature dependence of the Gribov parameter on the estimation of the heavy quark diffusion coefficient.   


\section{Numerical estimation of the diffusion coefficient}\label{results}
%

\begin{figure}[]
\center
\includegraphics[width=\linewidth]{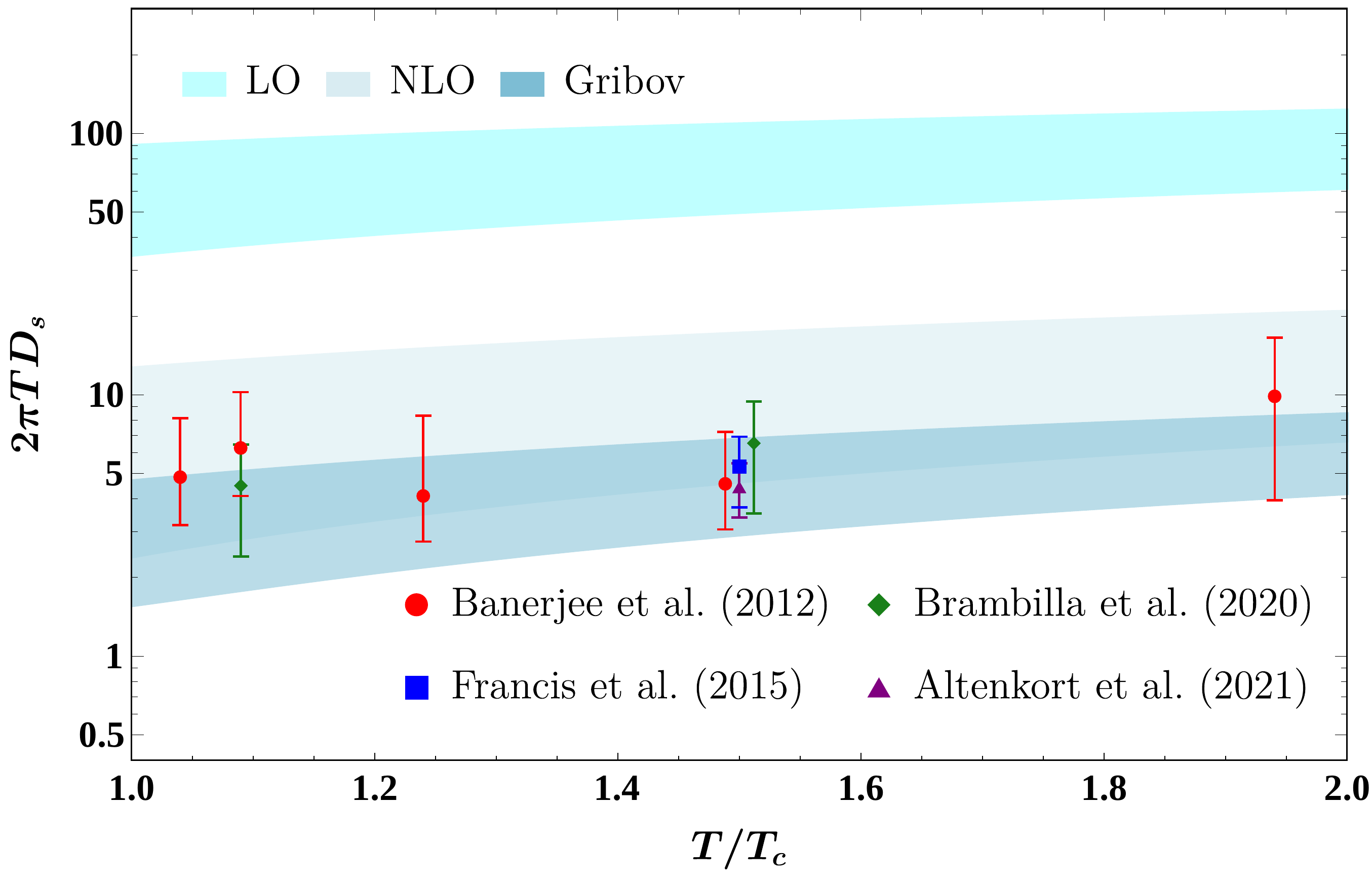}
\caption{The spatial diffusion coefficient  $(2\pi T)D_s$  as a function of the scaled temperature is compared with the lattice data from Refs.~\cite{Banerjee:2011ra,Francis:2015daa,Brambilla:2020siz,Altenkort:2020fgs}. At $T=1.5T_c$, a horizontal shift is introduced in the data for visual clarity. The bands refer to the variation of the scale parameter from $w=1$ to 4 and   $w_{\rm IR}$ to $w_{\rm UV}$ for the LO/NLO and the Gribov case, respectively.}
\label{fig:Ds}
\end{figure} 

To estimate the temperature dependence of the diffusion coefficient in the GZ framework, one has to  perform the momentum integral in Eq.~\eqref{q-integrated} numerically. Let us first consider the pure gluonic background. In that case, the non-trivial temperature dependence of the Gribov mass parameter (see Fig. \ref{gammaG_T_dependence}) is obtained by numerically solving  Eq.~\eqref{gap_eqn} along with the parametrized running coupling given in Eq.~\eqref{coupling}. The   scaled spatial diffusion coefficient, $(2\pi T)D_s$, can be obtained from the momentum diffusion coefficient ($\kappa$) by using the Einstein relation $(2\pi T)D_s=(4\pi T^3)/\kappa$.

\begin{figure}[]
\center
\includegraphics[width=\linewidth]{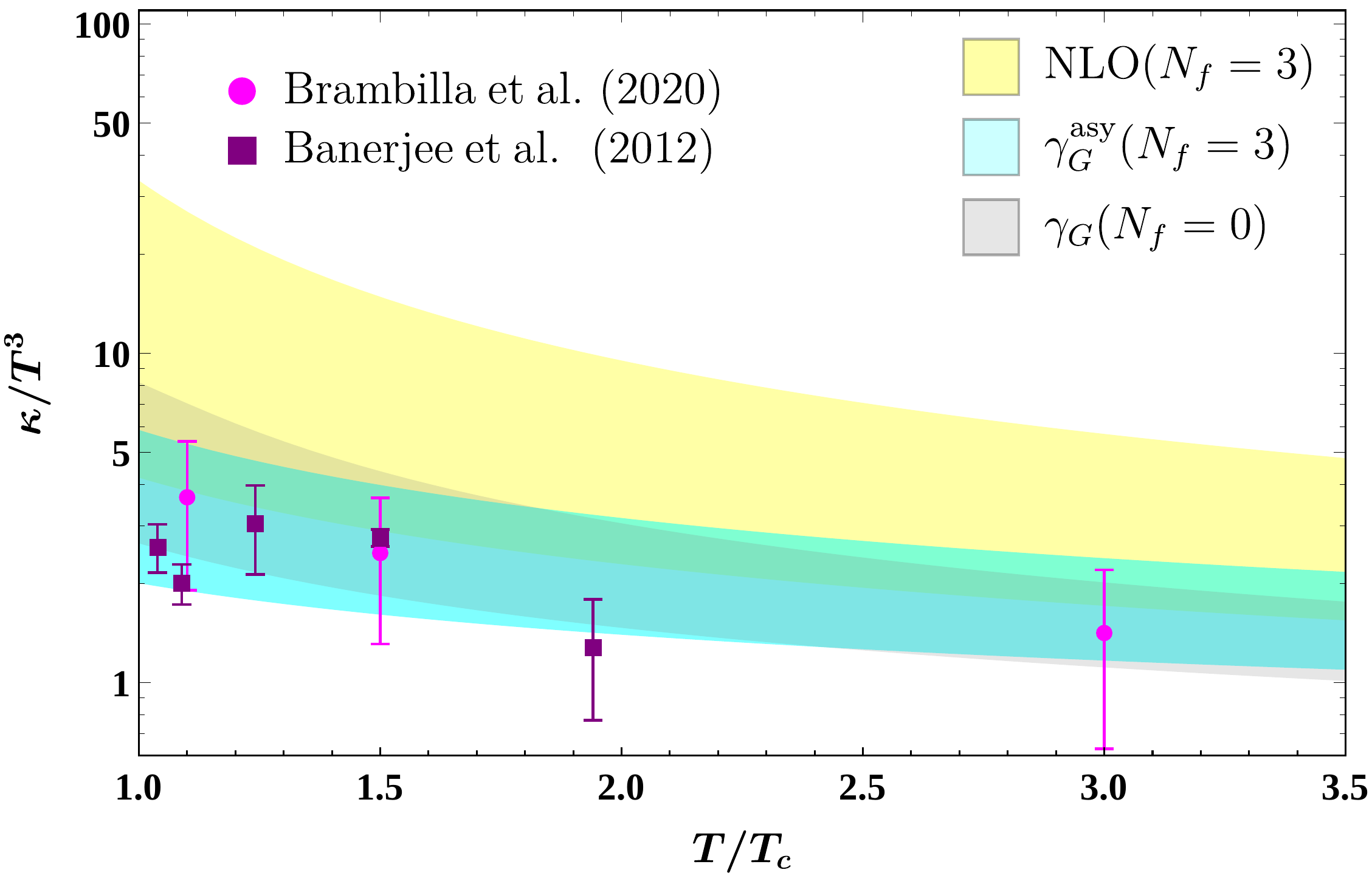}
\caption{The scaled momentum diffusion coefficient  with the asymptotic Gribov parameter is compared  with the NLO result considering  the 1-loop perturbative running coupling with $N_f=3$.  The bands refer to the variation of the scale ($w\,\pi T$) from $w=1$ to 4. The Gribov case with $N_f=0$ and the lattice data from~\cite{Banerjee:2011ra} and~\cite{Brambilla:2020siz}  are also shown for comparison.     
}
\label{fig:kappa_final_AM}
\end{figure} 

 The variation of the obtained $(2\pi T)D_s$ as a function of the scaled temperature $T/T_c$ is shown in Fig.~\ref{fig:Ds} along with the lattice estimations from Refs.~\cite{Banerjee:2011ra,Francis:2015daa,Brambilla:2020siz,Altenkort:2020fgs}. The  uncertainty  band, in this case, refers to the variation of the scale ($w T$) obtained by varying the  fit parameter $w$ in the range $w_{\rm IR}$ to $w_{\rm UV}$.  The corresponding values of the Gribov parameter (as shown in Fig.\ref{gammaG_T_dependence}) in each of the cases are obtained by solving the gap equation given in Eq. \eqref{gap_eqn}. It should be mentioned here that because of this variation of the Gribov parameter with $w$, the error band in the diffusion coefficient deviates from the trivial uncertainty expected from the overall multiplicative factor  (see, for example,  Eq. \eqref{q-integrated} with the overall $g^4$ factor). Similar reasoning also applies to the known LO and NLO expressions where the Debye mass $m_D$ varies with the parameter $w$. 
 As can be noticed from Fig.~\ref{fig:Ds}, the  temperature dependence of the Gribov-modified diffusion coefficient shows a reasonable agreement  with the  lattice estimations  within the uncertainties. Moreover, the linearly  increasing nature of the scaled spatial diffusion coefficient is consistent with the  data-driven  approach~\cite{Xu:2017obm,Li:2019lex} where a linearly increasing parametrization is usually implemented for the non-perturbative soft-part to compare the model output to the experimental data of the heavy-meson nuclear modification factor $R_{AA}$ and the elliptic flow $v_2$. As shown in Fig.~\ref{fig:Ds}, the Gribov-modified approach  significantly corrected the LO result. However, the behaviour of the Gribov improved spatial diffusion coefficient is qualitatively  similar to the result obtained incorporating the NLO correction, which is also shown in Fig. \ref{fig:Ds} for comparison. In this case, the approximate expression for the two-loop perturbative running~\cite{ParticleDataGroup:2012pjm}  is used in Eq.~\eqref{NLO} along with $T_c/\Lambda_{\rm \overline{MS} }=1.15$~\cite{Gupta:2000hr} and the uncertainty band is obtained by varying the energy scale symmetrically around $2\pi T$ by a factor of two. An overlap between the uncertainty bands of the NLO result and the Gribov-modified diffusion coefficient estimation  can be observed throughout the considered range of temperatures. Nevertheless, one can notice  a relatively narrower  uncertainty band in the GZ case, which essentially stems from the uncertainties in the lattice determination of the running coupling at finite temperatures~\cite{Kaczmarek:2004gv}.   

It is also interesting to compare the NLO result and the Gribov modified estimation, including the quark sector contribution. For this purpose, we consider the perturbative one loop running coupling given in Eq.~\eqref{per_coupling} (with the energy scale ($w\,\pi T$) varied considering $1\le w\le 4$)  along with the simple asymptotic temperature dependence  of the Gribov parameter (proportional to the chromo-magnetic scale) as given in Eq.~\eqref{gammaG_asym}.  Note that the  asymptotic temperature dependence of the Gribov parameter  has been considered earlier in the studies of the fermion dispersion relation~\cite{Su:2014rma} and the dilepton production rate~\cite{Bandyopadhyay:2015wua} in the GZ framework. In the present case, we incorporate this asymptotic form of the Gribov parameter in Eq. \eqref{q-integrated} to obtain the corresponding temperature dependence of the scaled momentum diffusion coefficient as shown in Fig~\ref{fig:kappa_final_AM}.  For comparison, the Gribov modified result with $N_f=0$ is also shown along with the pure glue lattice data  from Refs.~\cite{Banerjee:2011ra,Brambilla:2020siz}. It can be observed that the asymptotic form of the Gribov parameter with $N_f=3$ gives rise to an almost similar temperature dependence as obtained by solving the gap equation with the parametrized coupling in the $N_f=0$ case. Here also, an overlap between the uncertainty bands corresponding to the NLO correction and the Gribov-modified results can be observed throughout the considered temperature range. However, concentrating on a fixed energy scale $2\pi T$ for both cases, one can infer that, close to the critical temperature, the Gribov modified analysis indicates a smaller momentum diffusion coefficient value than the NLO estimation.


\section{Summary and outlook}\label{summary}
In this work, the heavy quark diffusion coefficient up to the leading order in $1/M$  expansion (where $M$ represents the mass of the heavy quark) has been studied in the phenomenologically relevant temperature regime using the Gribov-Zwanziger prescription. The relevant t-channel matrix amplitude at low momentum transfer has been obtained with the  IR-suppressed gluon propagator that depends on the Gribov parameter. For pure gluonic medium, the temperature dependence of this parameter has been obtained by solving the gap equation with a parametrized running coupling adopted from Ref.~\cite{Fukushima:2013xsa}. On the other hand, for $N_f=3$, we consider the asymptotic temperature dependence of the Gribov parameter, which is proportional to the chromo-magnetic scale. The temperature dependencies so obtained have been implemented in the estimation of the momentum diffusion coefficient based on the kinetic theory framework, and the corresponding spatial diffusion coefficient is obtained using the Einstein relation. We find that the Gribov-modified prescription significantly improves the LO estimation of the spatial diffusion coefficient resulting in a linearly increasing temperature dependence  consistent with the lattice estimations as well as the data-driven approaches. For both $N_f=0$ and $N_f=3$ cases, the estimated diffusion coefficients in the Gribov approach and the NLO perturbative approach overlap within the uncertainties (though with  reduced  uncertainties in the former approach).  Considering the median of the uncertainty bands with $N_f=3$, we find that, near the transition temperature, the GZ approach  results in  a smaller  momentum diffusion coefficient (closer to the value obtained in the pure glue scenario) compared to the NLO estimation. 

Note that the present study implements the original GZ approach for the simplicity of incorporating finite temperature modifications. A similar  analysis  based on the refined GZ approach~\cite{Canfora:2015yia} serves as an interesting future direction. Nevertheless, the significant improvement over the LO estimation of the diffusion coefficient obtained here (even with the asymptotic temperature dependence of the Gribov parameter) definitely encourages one to investigate the importance of the (chromo)magnetic scale in the estimation of other relevant transport properties of the medium~\cite{Florkowski:2015dmm,Jaiswal:2020qmj} as well as in the transport studies in the presence of non-trivial background~\cite{Fukushima:2015wck,Bandyopadhyay:2021zlm}.

\medskip

\section*{Acknowledgements}
%
The authors acknowledge Santosh Kumar Das for the fruitful discussion during the initial stages of the work.  A.B. acknowledges the support from the Alexander von Humboldt Foundation postdoctoral research fellowship in Germany. N.~H. is supported in part by the SERB-MATRICS under Grant No. MTR/2021/000939.



\bibliography{HQ}

\end{document}